\documentclass[10pt,conference]{./IEEEtran}
\usepackage{amsmath,amssymb,pst-all}
\usepackage{epsfig,subfigure,color,graphicx}

\newtheorem{lemma}{Lemma}
\newtheorem{theorem}{Theorem}
\newtheorem{corollary}{Corollary}
\newtheorem{definition}{Definition}

\renewcommand{\baselinestretch}{0.94}
\begin{document}

\title{ Bounds on Codes Based on Graph Theory }


\author{
\authorblockN{Salim Y. El Rouayheb}
\authorblockA{ECE Department \\
Texas A\&M University \\
College Station, TX 77843 \\
salim@ece.tamu.edu} \and
\authorblockN{Costas N. Georghiades}
\authorblockA{ECE Department \\
Texas A\&M University \\
College Station, TX 77843 \\
georghiades@ece.tamu.edu} \and
\authorblockN{Emina Soljanin}
\authorblockA{Math. Sc. Center \\
Bell Labs, Lucent \\
Murray Hill, NJ 07974 \\
emina@lucent.com}\and
\authorblockN{Alex Sprintson}
\authorblockA{ECE Department \\
Texas A\&M University \\
College Station, TX 77843 \\
spalex@ece.tamu.edu}
 }

\maketitle
\begin{abstract}

Let $A_q(n,d)$ be the maximum order (maximum number of codewords) of
a $q$-ary code of length $n$ and Hamming distance at least $d$. And
let $A(n,d,w)$ that of a binary code of constant weight $w$.
Building on results from algebraic graph theory and
Erd\H{o}s-ko-Rado like theorems in extremal combinatorics, we show
how several known bounds on $A_q(n,d)$ and $A(n,d,w)$ can be easily
obtained in a single framework. For instance, both the Hamming and
Singleton bounds can derived as an application of a property
relating the clique number and the independence number of vertex
transitive graphs. Using the same techniques, we also derive some
new bounds and present some additional applications.
\end{abstract}

\section{Introduction}
Let $\Sigma=\{0,1,\dots,q-1\}$ be an alphabet of order $q$. A
$q$-ary code $C$  of length $n$ and order $|C|$  is a subset of
$\Sigma^n$ containing $|C|$ elements (codewords). The weight
$wt(c)$ of a codeword $c$ is the number of its non-zero
entries. A $w$ constant weight code is a code where all the
codewords have the same weight $w$. The Hamming distance
$d(c,c')$ between two codewords $c$ and $c'$ is the number of
positions where they have different entries. The minimum
Hamming distance of a code $C$ is the largest integer $\Delta$
such that $\forall c,c'\in C, d(c,c')\geq \Delta$.

Let $A_q(n,d)$  be the maximal number of codewords that  a $q$-ary
code of length $n$ and minimum Hamming distance $d$ can possibly
contain (\cite[Chapter 17]{McWilliams}). $A(n,d,w)$ is defined
similarly for  binary codes with constant weight $w$. Finding the
values of  $A_q(n,d)$ and $A(n,d,w)$ is a basic problem in
``classical" coding theory \cite{Sloane,McWilliams}.

Finding a general exact expression for the maximal order of
codes is a difficult task. In fact, it was described in
\cite{Ahlswede2}, as ``a hopeless task". For this reason, much
of the research done has focused on bounding these quantities.

The dual problem, consisting of finding the maximal order of a set
of codewords satisfying an upper bound on their pairwise Hamming
distance (anticodes), is well studied in extremal combinatorics.
Surprisingly enough, it has a closed form solution \cite{Ahlswede,
Ahlswede2, Frankl}.

Using tools from algebraic graph theory, we draw a link between the
maximal order of codes and that of anti-codes. Then using results
like the celebrated Erd\H{o}s-ko-Rado theorem, we rederive some
known inequalities on $A_q(n,d)$ and $A(n,d,w)$ and other similarly
defined quantities and give some new bounds.

This paper is organized as follows. In Section \ref{graphintro}
we briefly introduce some of the needed background in graph
theory. In Section \ref{secbound} we show how the tools
introduced can be used to derive upper bounds on $A_q(n,d)$. In
Sections \ref{secconstant} and \ref{secdouble} we derive bounds
 on the maximal size of constant and doubly
constant weight codes, respectively. In Section \ref{secapp},
we show how the described techniques can be used to solve other
problems. We conclude in Section \ref{secconc}, where we
summarize our results and present some open questions.

\section{Graph Theory Background\label{graphintro}}

We start by giving a brief summary of some graph theoretical
concepts and results that will be  needed in this paper. For more
details, we refer the interested reader to \cite{Diestel} and
\cite{Godsil}.

Let $G(V,E)$ be an undirected graph, where $V$ is its vertex set and
$E$ is its edge set ($E\subseteq V\times V$). We also use
$\mathrm{V}(G)$ to denote the vertex set of $G$ and $\mathrm{E}(G)$
its edge set.
If $\{u,v\}$ is an edge in
$G$, i.e.\ $\{u,v\}\in E(G)$, we say that the vertices $u$ and $v$
are adjacent and write $u \sim v$.

The complement of a  graph $G$ is the graph $\bar{G}$ defined over
the same vertex set but where two vertices are adjacent in $\bar{G}$
iff they are not in $G$. We denote by $\omega(G)$ the \emph{clique
number} of a graph $G$, defined as  the largest number of vertices
of $G$ that are pairwise adjacent. In contrast  $\alpha(G)$, the
\emph{independence number} of $G$, is  the largest number of
vertices in $G$ such that no two of them are adjacent. It can be
easily seen that $\alpha(G)=\omega(\bar{G})$. In addition, the
\emph{chromatic number} $\chi(G)$ of $G$ is the minimum number of
colors needed to color its vertices such that different colors are
assigned to adjacent vertices.

\begin{definition}[Graph Automorphism  \cite{Godsil}]
Let $G(V,E)$ be a graph and $\phi$ a bijection from $V$ to
itself. $\phi$ is called an \emph{automorphism} of $G$ iff
$$\forall u,v\in V, u\sim v \Leftrightarrow \phi(u)\sim \phi(v).$$
\end{definition}\vspace{1.5mm}
The set of all automorphisms of $G$ is a group under composition; it
is called the automorphism group of $G$ and it is denoted Aut$(G)$.
For example, the complete graph on $n$ vertices $K_n$ has $S_n$, the
symmetric group of order $n$, as its automorphism group. In other
words, Aut$(K_n)\cong S_n$.

\begin{definition}[Vertex Transitive Graph \cite{Godsil}]
We say that graph $G(V,E)$ is vertex transitive iff
$$\forall u,v\in V, \exists \phi \in \text{Aut}(G)\text{ s.t. }\phi(u)=v.$$

\end{definition}\vspace{1.5mm}

\begin{definition}[Cayley Graphs]
Let $H$ be a group and $S\subset H$ such that $S$ is closed under
inversion and the identity element of $H$ $1_H \notin S$. The Cayley
graph $\mathcal{C}(H,S)$ is the graph with vertex set $H$ and where
for any $g,h\in H$, $g \sim h$ iff $hg^{-1}\in S$.
\end{definition}\vspace{1.5mm}

Next, we give without a proof an important result from \cite{Godsil}
(Lemma 7.2.2) that will be instrumental in deriving our results.

\begin{theorem}\label{corTrans}
Let $G(V,E)$ be a vertex transitive graph, then
$$\alpha(G)\omega(G)\leq |V(G)|.$$
\end{theorem}

\section{Bounds on Codes\label{secbound}}
\begin{definition}[Hamming Graph \cite{Sloane}]
The Hamming graph $H_q(n,d)$, $n\in
\mathbb{N}$ and $1\leq d\leq n$, has as vertices  all the $q$-ary
sequences of length $n$, and two vertices are adjacent iff
their Hamming distance is larger or equal to $d$. That is,
$V(H_q(n,d))=\Sigma^n$, where $\Sigma=\{0,1,\dots,q-1\}$.
and $u\sim v$ iff $d(u,v)\geq d$.
\end{definition}\vspace{1.5mm}

Recall that $A_q(n,d)$  denotes the maximum number of codewords in a
q-ary code of length $n$ and minimum Hamming distance $d$. When the
subscript is omitted we assume $q=2$, i.e. $A(n,d)=A_2(n,d)$. It can
be easily seen that $A_q(n,d)=\omega(H_q(n,d))$.

Let $S_{n,d}$, $1\leq d\leq n$, be a subset of the group
$(\mathbb{Z}_q^n,+)$, where addition is done modulo $q$, such that
$S_{n,d}=\{s\in \mathbb{Z}_q^n; wt(s)\geq d \}$. It is easy to check
that $S_{n,d}$ is closed under inversion and does not contain the
identity element (the all zero sequence). The next lemma asserts
that the Hamming graph is in fact a Cayley graph.
\begin{lemma}\label{HammingCayley}
$H_q(n,d)=\mathcal{C}(\mathbb{Z}_q^n,S_{n,d}).$
\end{lemma}\vspace{1.5mm}
\begin{proof}
Take $\Sigma=(\mathbb{Z}_q,+)$. The result then follows easily from
the fact that $\forall x,y \in \mathbb{Z}_q^n, $ $d(x,y)=wt(x-y)$.
\end{proof}

\begin{lemma}\label{transhamming}
The Hamming graph $H_q(n,d)$ is vertex transitive.
\end{lemma}\vspace{1.5mm}
\begin{proof}
Follows From Lemma \ref{HammingCayley} and the fact that Cayley
Graphs are vertex transitive \cite[Thm.~3.1.2]{Godsil}.

For a clearer presentation, we also give here a direct proof.
 Take $\Sigma=(\mathbb{Z}_q,+)$.
And $\forall u,v,x\in \Sigma^n$, define the function
$\phi_{u,v}(x)=x+v-u$. $\phi_{u,v}(x)$ is an automorphism of
$H_q(n,d)$. In fact,
$d(\phi_{u,v}(x),\phi_{u,v}(y))=d(x+v-u,y+v-u)=\text{wt}(x+v-u
-(y+v-u))=\text{wt}(x-y)=d(x,y)$. Also, $\phi_{u,v}(x)$ takes $u$ to
$v$.
\end{proof}

\begin{corollary}\label{hammingineq}
$ A_q(n,d) \alpha(H_q(n,d))\leq q^n$
\end{corollary}
\vspace{1mm}
\begin{proof}
Follows from Lemma  \ref{transhamming} and Thm.~\ref{corTrans}.
\end{proof}

Notice that $\alpha(H_q(n,d))$, the independence number of the
Hamming graph $H_q(n,d)$, is actually the maximum number of
sequences such that the Hamming distance between any two of them is
at most $d-1$. Following \cite{Ahlswede}, we define  $N_q(n,s)$ to
be the maximum number of $q$-ary sequences of length $n$ that
intersect pairwise (have the same entries) in at least $s$
positions. It follows that
\begin{equation}\label{equInd}
    \alpha(H_q(n,d))=N_q(n,t); \quad  \text{with }t=n-d+1
\end{equation}

\begin{lemma}[Singleton Bound]\label{lemSingleton}
$A_q(n,d)\leq q^{n-d+1}$
\end{lemma}\vspace{1.5mm}
\begin{proof}
Consider the set $T(n,t)$ of $q$-ary sequences of length $n$ that
all have the same element in the first $t=n-d+1$ entries. By
definition, $N_q(n,t)\geq |T(n,t)|= q^{n-t}$. Then, by
(\ref{equInd}) and Corollary \ref{hammingineq}, $A_q(n,d)\leq
\frac{q^n}{q^{n-t}}=q^{n-d+1}$.
\end{proof}

\begin{lemma}[Hamming Bound]\label{lemHamming}
\[
    A_q(n,d)\leq
    \frac{q^n}{\sum_{i=0}^{\lfloor\frac{d-1}{2}\rfloor}\binom{n}{i}(q-1)^i}.
\]
\end{lemma}\vspace{1.5mm}
\begin{proof}The proof is similar to that of Lemma~\ref{lemSingleton} and is done
by finding a different lower bound on $N_q(n,t)$. In fact,
consider the ball  $B(n,r)=\{x\in\Sigma^n;wt(x)\leq r \}$. By
the triangle inequality,  $\forall x,y\in B(n,\lfloor
\frac{d-1}{2}\rfloor), d(x,y)\leq d-1$. Therefore $N_q(n,t)\geq
|B(n,\lfloor \frac{d-1}{2}\rfloor)|$, and
$A_q(n,d)\leq\frac{q^n}{ B(n,\lfloor \frac{d-1}{2}\rfloor)}$.
\end{proof}

The number $N_q(n,t)$ is well studied in extremal combinatorics
\cite{Ahlswede} \cite{Frankl}, and a closed form for it is
known. Thus, exact expressions of $N_q(n,t)$ can be used to
derive better upper bounds on $A_q(n,d)$. For instance, if
$n-t$ is even, $N_2(n,t)=\sum_{i=0}^{\frac{n-t}{2}}
\binom{n}{i}$. Thus, in this case, $B(n,\lfloor
\frac{d-1}{2}\rfloor)$ is a maximal anticode. However, when
$n-t$ is odd, $N_2(n,t)= 2 \sum_{i=0}^{\frac{n-t-1}{2}}
\binom{n-1}{i}$ \cite[Thm.~Kl]{Ahlswede} and
\cite{Kleitman}. Therefore, we obtain the following lemma.

\begin{lemma}\label{firstbound}
\begin{equation}\label{binaryb}
    A(n,d)\leq
    \frac{2^{n-1}}{\sum_{i=0}^{\frac{d-2}{2}}\binom{n-1}{i}}, \quad \text{if $d$ is even}.
\end{equation}
\end{lemma}\vspace{1.5mm}

Notice that the above bound is tighter than the Hamming bound for even $d$ since
$$ 2 \sum_{i=0}^{\frac{d-2}{2}} \binom{n-1}{i}
-\sum_{i=0}^{\frac{d-2}{2}} \binom{n}{i}
=\binom{n-1}{\frac{d-2}{2}}>0.$$
This new improved Hamming
bound was recently proven in \cite{Matsumoto} using different
techniques than the one presented here.

Next we give a new upper bound on $A_q(n,d)$ for alphabets of
arbitrary size.
\begin{lemma}
For $q\geq 3$, $t=n-d+1$ and $r=\lfloor \min\{\frac{n-t}{2},
\frac{t-1}{q-2}\}\rfloor$,

\begin{equation}\label{qarybound1}
 A_q(n,d)\leq\frac{q^{t+2r}}{\sum_{i=0}^{r}\binom{t+2r}{i}(q-1)^i}.
\end{equation}
\end{lemma}\vspace{1.5mm}

\begin{proof}
The proof follows from Corollary \ref{hammingineq} and Thm.~2 in
\cite{Frankl} or the Diametric Theorem of \cite{Ahlswede}.
\end{proof}

Note that for $q\geq t+1$, $N_q(n,t)=q^{n-t}$ \cite[Corollary
1]{Frankl}, i.e. a maximal anticode would be the trivial set
$T(n,t)$ described in the proof of Lemma~\ref{lemSingleton}. In
this case, the bound of (\ref{qarybound1}) boils down to the
Singleton bound.

For $d$ even and $n$ not much larger than $t$, the next lemma
provides an improvement on the Hamming bound for nonbinary
alphabets.

\begin{lemma}\label{lemHamming2}
For $d$ odd and $n\leq t+1+\frac{\log t}{\log (q-1)}$
\begin{equation}
A_q(n,d)\leq \frac{q^{n-1}}{\sum_{i=0}^{\frac{d-2}{2}}
\binom{n-1}{i}(q-1)^i}
\end{equation}
\end{lemma}\vspace{1.5mm}
\begin{proof}
Under the conditions of this lemma,
$N_q(n,t)=q\sum_{i=0}^{\frac{d-2}{2}} \binom{n-1}{i}(q-1)^i$
\cite[Eq.\ 1.7]{Ahlswede}. The result then follows from  Corollary
\ref{hammingineq}.
\end{proof}

\section{Bounds for Constant Weight Codes\label{secconstant}}

Let $A(n,2\delta,w)$ be the maximum possible number of
codewords in a  \emph{binary} code of  length $n$, constant
weight $w$ and minimum distance $2\delta$ \cite{Sloane, Vardy}.

Define the graph $K(n,2\delta,w)$ as the graph whose vertices
are all the binary sequences of length $n$ and weight $w$ and
where two vertices $u,v$ are adjacent iff $d(u,v)\geq 2\delta$.
It can be easily seen that
$A(n,2\delta,w)=\omega(K(n,2\delta,w))$.

Let $\binom{[n]}{w}$ denote the set of all subsets of
$[n]=\{1,2,\dots,n\}$ of order $w$. There is a natural bijection
$\nu$ between $\mathrm{V}(K(n,2\delta,w))$ and  $\binom{[n]}{w}$.
Namely, $\forall u \in \mathrm{V}(K(n,2\delta,w))$, $\nu(u)=U=\{i;
u(i)=1\}$.

\begin{lemma}\label{lemequivalence}
$\forall p,q\in  \mathrm{V}(K(n,2\delta,w)), p\sim q$ iff $|P\cap
Q|\leq w-\delta$ where $P=\nu(q)$ and $Q=\nu(q)$.
\end{lemma}\vspace{1.5mm}
\begin{proof}
$2\delta\leq d(p,q)=|(P\cap \bar{Q})\cup (\bar{P}\cap Q)|=2w-2|P\cap
Q|.$
\end{proof}

\begin{lemma}\label{lemXtrans}
$K(n,2\delta,w)$ is vertex transitive.
\end{lemma}\vspace{1.5mm}
\begin{proof}
For any two vertices $p,q$ of $K$, any bijection on $[n]$ such that
the image of $P=\nu(p)$ is $Q=\nu(q)$, takes $p$ to $q$ and belongs
to $Aut(K)$.
\end{proof}

The first result that follows directly  from Lemma \ref{lemXtrans}
is the Bassalygo-Elias inequality \cite{Vardy}. We first recall some
additional  results in graph theory.
\begin{definition}[Graph Homomorphism]
Let $X$ and $Y$ be two graphs. A mapping $f$ from $\mathrm{V}(X)$ to
$\mathrm{V}(Y)$ is a homomorphism if $\forall x,y\in \mathrm{V}(X)$
 $x\sim y \Rightarrow f(x)\sim f(y).$
\end{definition}

\begin{theorem}\label{lem7}
If $Y$ is vertex transitive and there is a homomorphism from $X$ to
$Y$, then
\[
    \frac{|V(X)|}{\alpha(X)}\leq  \frac{|V(Y)|}{\alpha(Y)}
\]
\end{theorem}
\begin{proof}
An application of Lemma 7.14.2 in \cite{Godsil}.
\end{proof}

\begin{lemma}[Bassalygo-Elias inequality]\label{lemElias}
\[
  A(n,d)\leq \frac{2^n}{\binom{n}{w}}A(n,d,w)
\]
\end{lemma}\vspace{1.5mm}
\begin{proof}
Consider the two graphs $Y=\bar{H}(n,d)$ and $X=\bar{K}(n,d,w)$. $Y$
is vertex transitive. Since $X$ is an induced subgraph of $Y$, the
inclusion map is a homomorphism that takes $X$ to $Y$. The result
then follows from applying Thm.~\ref{lem7}.
\end{proof}

By the same token, we can show the below equalities

\begin{lemma}\label{inequalities}
\begin{align}
\label{mybound} A(n,d,w)&\leq \frac{n-w+1}{w} A(n,d+2,w-1)\\
\label{mybound2}A(n,d,w)&\leq \frac{n+1}{w+1}A(n+1,d+2,w+1)\\
   \label{johnson1} A(n,d,w)&\leq \frac{n}{w}A(n-1,d,w-1)\\
   \label{johnson2} A(n,d,w)&\leq \frac{n}{n-w} A(n-1,d,w)
    \end{align}
\end{lemma}\vspace{1.5mm}
\begin{proof}
We start by proving inequality \ref{mybound}. Let $\phi$ be a
mapping from $\binom{[n]}{w-1}$ to $\binom{[n]}{w}$, such that
$\forall P\in \binom{[n]}{w-1}, P\subset\phi(P)$. $\phi$ is a
homomorphism from $K(n,d+2,w-1)$ to $K(n,d,w)$. In fact, $\forall
P,Q\in K(n,d+2,w-1)$ such that $P\sim Q, |\phi(P)\cap\phi(Q)|\leq
|P\cap Q|+2\leq w-1-(d+2)/2+2=w-d/2$ (by Lemma
\ref{lemequivalence}). Therefore, $\phi(P)\sim\phi(Q)$. The
inequality then follows by applying Thm.~\ref{lem7}.

To prove  inequality \ref{mybound2}, take the homomorphism $\phi$
from $K(n+1,d+2,w+1)$ to $K(n,d,w)$ to be
$\phi(X)=X\setminus\{\max_{x\in X} x\}, \forall X\in
\binom{[n+1]}{w+1}$.

The rest of the inequalities can be proved similarly by considering
the  corresponding graphs and taking the homomorphism to be the
inclusion map.
\end{proof}

The first two inequalities are new, whereas inequalities
\ref{johnson1} and \ref{johnson2} were first proven by Johnson in
\cite{Johnson}.

Similarly, we can show the following inequalities regarding
$A_q(n,d)$.

\begin{lemma}\label{inequalities2}
\begin{align*}
    A_q(n,d)&\leq \frac{1}{q} A_q(n+1,d+1)\\
    A_q(n,d)&\leq q A_q(n-1,d)\\
    A_q(n,d)&\leq \frac{q^n}{(q-1)^n} A_{q-1}(n,d,w)
\end{align*}
\end{lemma}\vspace{1.5mm}

\begin{lemma}\label{thconst}
Let $t=w-\delta+1$.
\begin{equation}\label{boundconst}
  A(n,2\delta,w)\leq  \frac{\binom{n}{w}}{\binom{n-t}{w-t}}
\end{equation}
\end{lemma}\vspace{1.5mm}

\begin{proof}
Let $G=K(n,d,w)$. Since $G$ is vertex transitive, we have
$$A(n,2\delta,w)\alpha(G)\leq |V(G)|=\binom{n}{w}.$$

Define $M(n,w,s)$ as in \cite{Ahlswede2} to  be the  maximum number
of subsets of $[n]$ of order $w$ that intersect pairwise in at least
$s$ elements. By Lemma \ref{lemequivalence}, $\alpha(G)=M(n,w,t)$.
But, $M(n,w,t)\geq \binom{n-t}{w-t}$ (for instance, consider the
system of all subsets of $[n]$ of order $w$ that contain the set
$\{1,2,\dots,t\}$).
\end{proof}

The bound of Lemma \ref{thconst} is actually the same as the one in
Thm.~12 in \cite{Vardy} which was given with a different proof.

One can improve on the bound of Lemma \ref{thconst} by using
the exact value of $M(n,w,t)$ \cite{Ahlswede2}. It is known
that for $n\geq (w-t+1)(t+1)$, $M(n,w,t)=\binom{n-t}{w-t}$
\cite{Erdos, Wilson}. However, this is not the case for lower
values of $n$.

\begin{lemma}
Let $t=w-\delta+1$ and $r=\max\{0, \lceil
\frac{\delta(w-\delta)}{n-d}-1 \rceil\}$, then

\begin{equation}\label{boundconst}
  A(n,2\delta,w)\leq
  \frac{\binom{n}{w}}{\sum_{i=t+r}^{w}\binom{t+2r}{i}\binom{n-t-2r}{w-i}};
\end{equation}
with $\binom{n}{k}=0$ when $k> n$.
\end{lemma}\vspace{1.5mm}
\begin{proof}(sketch)
$A(n,d,w)\leq \frac{\binom{n}{w}}{M(n,w,t)}$, then use the exact
value of $M(n,w,t)$ given by the main theorem of \cite{Ahlswede2}.
\end{proof}

\section{Bounds for Doubly Bounded Weight Codes\label{secdouble}}

Let $T(w_1,n_1,w_2,n_2,d)$ be the maximum number of codewords in a
doubly constant weight binary code of minimum distance $d$, length
$n=n_1+n_2$ and constant weight $w=w_1+w_2$,  where the first $n_1$
entries of each codewords have exactly $w_1$ ones \cite{Lev}.
$T'(w_1,n_1,w_2,n_2,d)$ is defined similarly but where the first
$n_1$ entries of each codewords have at most $w_1$ ones
\cite{Vardy}.

\begin{lemma}
\begin{align}
\label{levineq}A(n,d,w)&\leq
\frac{\binom{n_1+n_2}{w_1+w_2}}{\binom{n_1}{w_1}\binom{n_2}{w_2}}T(w_1,n_1,w_2,n_2,d)\\
\label{myineq}A(n,d)&\leq
\frac{2^n}{\sum_{i=0}^{w_1}\binom{n_1}{i}\binom{n_2}{w_1+w_2-i}}T'(w_1,n_1,w_2,n_2,d)
\end{align}
\end{lemma}\vspace{1.5mm}
\begin{proof}
Same as Lemma \ref{lemElias}.
\end{proof}

Note that inequality (\ref{levineq}) was first proven in
\cite{Lev}, whereas inequality (\ref{myineq}) is new. Several
other bounds on $T(w_1,n_1,w_2,n_2,d)$ known in literature,
such as Theorem~36 in \cite{Vardy}, can be also easily
obtained in the same way. The next lemma establishes some
additional new bounds. \small

\begin{lemma}\vspace{-3mm}
\begin{align*}
T(w_1,n_1,w_2,n_2,d)&\leq \binom{n_2}{w_2}A(n_1,w_1,d-2w_2)
    \text{ if }
    d-2w_2\geq 0\\
    T(w_1,n_1,w_2,n_2,d)&\leq \binom{n_1}{w_1}A(n_2,w_2,d-2w_1)
    \text{ if }
    d-2w_1\geq 0\\
    T(w_1,n_1,w_2,n_2, d)&\leq \frac{n_1-w_1+1}{w_1} T(w_1-1,n_1,w_2,n_2, d+2)\\
T(w_1,n_1,w_2,n_2, d)&\leq
\frac{n_1+1}{w_1+1}T(w_1+1,n_1+1,w_2,n_2, d+2)\\
     T(w_1,n_1,w_2,n_2, d)&\leq \frac{n_2-w_2+1}{w_2} T(w_1,n_1,w_2-1,n_2, d+2)\\
T(w_1,n_1,w_2,n_2, d)&\leq
\frac{n_2+1}{w_2+1}T(w_1,n_1,w_2+1,n_2+1, d+2)
    \end{align*}
\end{lemma}\vspace{1.5mm}
\normalsize

\section{Other Applications}\label{secapp}

In this section we demonstrate how the above techniques can be
helpful in solving other problems. For instance, we show how to
compute $N_q(n,1)$, the maximum number of $q$-ary sequences of
length $n$ intersecting pairwise in at least one position
\cite{Ahlswede}.

\begin{lemma}\label{lemberge}
$N_q(n,1)=q^{n-1}$
\end{lemma}\vspace{1.5mm}
\begin{proof}
Let $G=H_q(n,n)$; $N_q(n,1)=\alpha(G)$. Now, consider the set
of $q$ sequences where the entries in the $i$-th sequence are
all the same and equal to $i$, hence $\omega(G)\geq q$. But
$\omega(G)\leq q$ since the first entries of all sequences in a
clique in $G$ should contain different letters. Therefore,
$\omega(G)=q$. By Lemma \ref{transhamming}, we get
$N_q(n,1)\leq q^{n-1}$. But $N_q(n,1)\geq q^{n-1}$(see the
proof of Lemma~\ref{lemSingleton}).
\end{proof}

The next lemma gives the chromatic number of certain Hamming graphs.
\begin{lemma}\label{lemchro}
$\chi(H_q(n,d))=q^{n-d+1}$, for $q\geq n-d+2$, $1\leq d \leq n$.
\end{lemma}\vspace{1.5mm}
\begin{proof}
From the definitions, it follows that for any graph $G$,
$\chi(G)\geq \frac{|\mathrm{V}(G)|}{\alpha(G)}.$ But,
$\alpha(H_q(n,d))=q^{d-1}$ \cite[Corollary 1]{Frankl}. Therefore,
$\chi(H_q(n,d))\geq\frac{q^n}{q^{d-1}}= q^{n-d+1}$.

Let $\phi$ be a mapping from $\Sigma^n$ to $\Sigma^{n-d+1}$
consisting of deleting the last $d-1$ entries of a sequence. $\phi$
is a homomorphism from $H_q(n,d)$ to $H_q(n-d+1,1)=K^{n-d+1}$, where
$K^\ell$ is the complete graph on $\ell$ vertices. Therefore,
$\chi(H_q(n-d+1,1))\leq \chi( K^{n-d+1})=q^{n-d+1}$ \cite[Lemma
1.4.1]{Godsil}.
\end{proof}

Let  $v(G)$ be the Lov\'{a}sz upper bound \cite{x} on the zero error
capacity $\Theta(G)$ \cite{Shannon} of a graph $G$. We recall the
following two results of \cite{x}.

\begin{lemma}\label{Lov1}

$\alpha(G)\leq \Theta(G)\leq v(G)$
\end{lemma}

\begin{theorem}\label{Lov2}
If $G(V,E)$ is vertex transitive then $v(G) v(\bar{G})=|V|$.
\end{theorem}
In the following, we give a partial answer to  a question
raised in the conclusion of \cite{x}, namely ``Find further
graphs with $v(G)=\Theta(G)$".

\begin{lemma}\label{lemvt1}
The following graphs satisfy $v(G)=\Theta(G)$
\begin{enumerate}
  \item $H_q(n,d)$ when there exists a q-ary perfect code of length $n$ and minimum
distance $d$.
\item $H_q(n,d)$ when $q\geq n-d+2$ and there exists a q-ary MDS code
of length $n$ and minimum distance $d$.
\item $H_q(n,n)$.
\end{enumerate}
\end{lemma}
\begin{proof}
Let $G$ be a vertex transitive  graph such that $\alpha
(G)\alpha(\bar{G})=|\mathrm{V}(G)|$. Then, applying Lemma
\ref{Lov1} to $G$ and $\bar{G}$ and multiplying the two
resulting equations we get
$\frac{\Theta(G)}{v(G)}=\frac{v(\bar{G})}{\Theta(\bar{G})}\geq
1$. Therefore, $\Theta(G)=v(G)$. One can check that the graphs
$G$ belonging to the three families mentioned above satisfy
$\alpha (G)\alpha(\bar{G})=|\mathrm{V}(G)|$.
\end{proof}

\section{Conclusion }\label{secconc}
We constructed vertex transitive graphs where a code corresponds to
a clique and an anti-code  to an independent set. Thus, we
established a connection between the maximal order of  codes and
that of anti-codes. Using intersection theorems for systems of
finite sets and that of  finite sequences, we  provided a framework
where several known bounds on code size follow easily and  new
inequalities can be derived.

Several questions naturally arise here.

\begin{enumerate}
  \item What are  the zero error capacities of the graphs $H$ and $K$ and their
complements $\bar{H}$ and $\bar{K}$? What are   the values of the
$v$ function of these graphs.  Note, that these quantities can be
useful to derive bounds for $A_q(n,d)$ and $A(n,d,w)$ using Lemma
\ref{Lov1} and Thm.~\ref{Lov2}.
  \item From a graph theoretical
standpoint, trying to extend the result of Lemma \ref{lemchro} by
finding the chromatic number of the above graphs is also an
interesting question, and can have  applications to coding theory
and cryptography.
\item Perfect codes are codes who achieve the Hamming bound. We
gave here many upper bounds lower than the Hamming bound in specific
cases (Lemma~\ref{firstbound}, (\ref{qarybound1}), Lemma~\ref{lemHamming2}
and (\ref{boundconst})); thus ruling out the existence of perfect
codes there. It is an interesting question to find whether there
exist "nearly perfect codes" that can achieve these new bounds.
\end{enumerate}

\end{document}